\documentclass[fleqn,10pt]{wlscirep}
\usepackage[utf8]{inputenc}
\usepackage[T1]{fontenc}
\title{Babinet-Complementary Structures for Implementation of  
Pseudospin-Polarized Waveguides}

\author[1,*]{Haddi Ahmadi}
\author[2,+]{Amin Khavasi}
\affil[ ]{Department of Electrical Engineering, Sharif University of Technology, Tehran 11155-4363, 
Iran}

\affil[*]{ahmadihaddi@gmail.com}

\affil[+]{khavasi@sharif.ir}

\begin{abstract}
In this work, we prove a theorem that states the electromagnetic (EM) duality correspondence
between eigenmodes of complementary structures, induces counterpropagating spin-polarized 
states in different types of waveguides where mirror reflection symmetries are preserved around 
one (or more) arbitrary plane(s). Similar to photonic topological insulators (PTIs), which support 
topologically non-trivial direction-dependent spin polarizations, our pseudospin-polarized 
systems support one-way states that manifest robustness, however, the advantage of our 
structures is that they can be implemented in extremely broad bandwidth simply using artificial 
dual impedance surfaces. Consequently, there is no need to bulk electromagnetic materials. On 
the basis of our theory, the concept of the pseudospin-polarized waveguide can be realized using 
Babinet complementary structures, ranging from microwave to THz regime. We design 
and develop various unidirectional waveguides and spin-filtered feature in the microwave regime 
is investigated. 
\end{abstract}
\begin{document}

\flushbottom
\maketitle
%
%
\thispagestyle{empty}

\section*{Introduction}
Waveguiding structures are considered as enabling components, transforming electromagnetic 
power and/or signal from one end to another with minimum conceivable loss of EM energy [1,2]. 
Recent progress in designing artificially patterned planar structures has led to the implementation 
of open waveguides incorporating complementary metasurfaces [3-5]. Metasurfaces as the 
planar counterparts of metamaterials benefit from the ease of fabrication and have the capability to 
manipulate surface waves propagation by shaping the sizes of their unit cells [6-9]. 
Metasurfaces are also suitable for integration with microwave and nanophotonic devices due to 
their planar constitution [10-12].\\
 Recently, there has been a demonstration of a new EM mode based 
On complementary impedance boundaries concept, called a line wave (LW) [3]. The interface of 
two complementary metasurfaces forms spin-polarized line mode, which is induced by imposing 
the inversion-symmetries. The LW exhibits robust unidirectional propagation and broad 
operating bandwidth. Similarly, symmetry-preserved pseudospin states have been observed in a 
channel with a perfect electric conductor (PEC) and effective perfect magnetic conductor (PMC) 
boundaries [13]. Nevertheless, the use of dual EM boundary conditions in the form of PEC and PMC 
limits the working bandwidth of a spin-polarized waveguide. On the other hand, time-reversal 
invariant PTIs have been proposed that exhibit pseudospin-polarized one-way transport [14-19]. The
protected propagation of interface states in these systems arises from energy bands with 
nontrivial topological features. Alternatively, photonic systems with broken time-reversal 
symmetry support one-way edge states confined to the interface between two magneto-optic crystals. These nonreciprocal structures, which are the precise counterparts of quantum Hall 
effect exhibit robustness against structural deformations as swell [20-23]. However, robust transport 
implementations on the basis of magneto-optic effects and PTIs, suffer from restricted bandwidth 
and inevitable design complexities [14,23]. On the other hand, the constitution of the pseudospin states 
in an arbitrary system whose boundary conditions are formed by complementary structures has
not been comprehensively studied. Here, it is shown that the protected transport characteristic
within an ultra-wide frequency range, can be achieved in a straightforward manner by imposing
complementary impedance boundary conditions.\\
In this paper, we prove that how the eigenstates of a system with EM complementary boundary 
conditions, form a pseudospin pair. Next, we design numerous broadband pseudospin-suppressed waveguides, and investigate the immune spin-dependent transportation. In 
pseudospin transport systems with complementary impedance boundaries, the spatial inversion 
symmetries, which is the outcome of EM duality between the eigenfields of complementary 
surfaces leads to forming polarization-momentum locked spin-up forward and spin-down 
backward states, which are robust against spatial perturbations. In such systems, disturbances do 
not establish backward propagating and they are strongly filtered owing to preserving time reversal symmetries. Our finding is universal and confirms the possibility to generalize the 
concept of the spin-polarized protected transport into all classical waveguides. We examine this theory
on some open and closed waveguides. The waveguides studied in this research are 
formed by complementary metasurfaces having isotropic sheet impedances. Our waveguides are suitable for applications in sensing, chiral quantum coupling, directional couplers, one-way system isolators, and ultra wide-band reconfigurable devices. This work opens a 
wide avenue for designing and developing different pseudospin-polarized waveguides.

\section*{Theorem of Pseudospin States}
There is an EM duality for the eigenstates on complementary metallic ultrathin structures 
inserted in a homogenous, isotropic, and source-free medium [24]. The explicit duality 
correspondence of EM fields on complementary formed thin metallic films warrants that the 
surface modes on dual-complementary structures share strictly the same dispersion. Indeed, 
this spatial correlation of EM fields is a connotation of Babinet principle [25].\\
Consider two 
interfaced complementary metasurfaces extended uniformly in the
$\pm y$ 
directions, as 
shown in Figure 1. Assuming that the complementary structures are suspended in free space coinciding with the x-y plane. At subwavelength frequencies a
circular patch metasurface and its Babinet complement, i.e. a circular mesh metasurface, exhibit a 
dominant mode of transverse-electric (TE) and transverse-magnetic (TM) waves, 
respectively. Let us consider that the TM surface comprises $ E_x, H_y,$ and $E_z$ field components at
any point $(x,-y, z)$. Hence, the field components of the complementary surface, i.e., $H_x, E_y,$ and 
$H_z$ at any point $(x,y,z)$ based on EM duality, can be written as
\begin{figure}[ht]
\centering
\includegraphics[width=\linewidth]{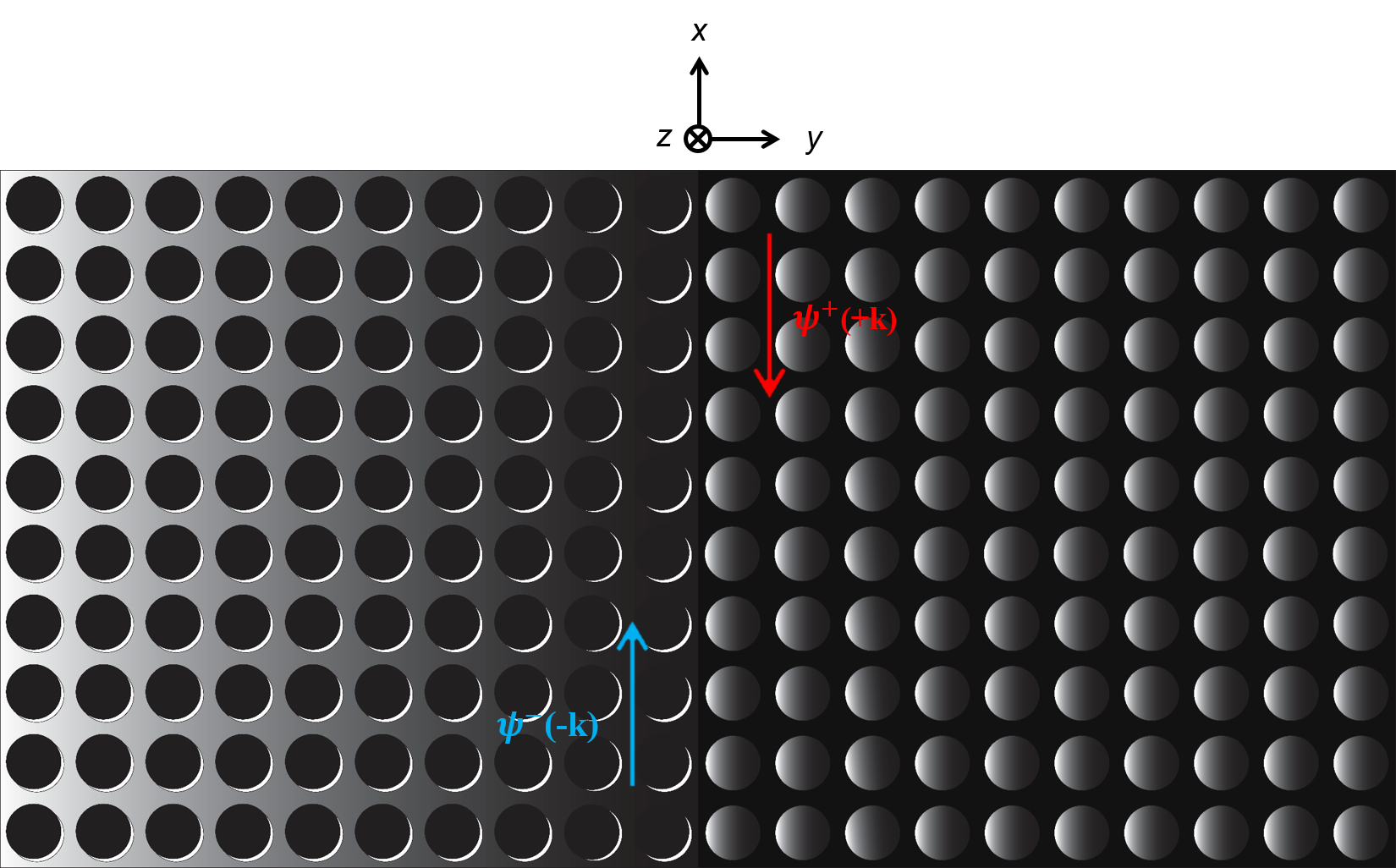}
\caption{ Two interfaced dual metasurfaces that support pseudospin states at their interface line. The complementary metasurfaces satisfy $\epsilon_r(x,-y,z)=\mu_r(x,y,z)$ and $\mu_r(x,-y,z)=\epsilon_r(x,y,z)$ symmetries because they constitute a pair of mirror images.}
\label{fig:Figure1}
\end{figure}

\begin{equation}
E_y(x,y,z)=\sqrt{\frac {\mu_0} {\epsilon_0}} H_y(x,-y,z)
\end{equation}
\begin{equation}
H_x,_z(x,y,z)=-\sqrt{\frac {\epsilon_0} {\mu_0}} \,E_x,_z(x,-y,z)
\end{equation}
Equation (1) describes the relationship between EM fields in pseudospin up state. According to 
the definition of the pseudospin up state,
$E_y$
field component at any point $(x,y,z)$ and 
$H_y$
field 
component at any arbitrary point $(x,-y,z)$ are in phase, while the
$H _x,_z$
field components and the 
$E_x,_z$
field components are out of phase.\\
The differential form of Maxwell's equations describing the propagation of TM surface wave at point 
$(x,-y,z)$ are given by

\begin{equation}
    -\frac {\partial{ }} {\partial{x}} E_z(x,-y,z)+\frac {\partial{ }} {\partial{z}} E_x(x,-y,z)=-j\omega\mu_0\mu_r(x,-y,z)H_y(x,-y,z)
\end{equation}
\begin{equation}
     \frac {\partial{ }} {\partial{x}} H_y(x,-y,z)=j\omega\epsilon_0\epsilon_r(x,-y,z)E_z(x,-y,z)
\end{equation}
\begin{equation}
    - \frac {\partial{ }} {\partial{z}} H_y(x,-y,z)=j\omega\epsilon_0\epsilon_r(x,-y,z)E_x(x,-y,z)
\end{equation}
Similarly, the differential form of Maxwell's equations for the complementary surface wave at point 
$(x,y,z)$ are given by
\begin{equation}
    -\frac {\partial{ }} {\partial{x}} H_z(x,y,z)+\frac {\partial{ }} {\partial{z}} H_x(x,y,z)=j\omega\epsilon_0\epsilon_r(x,y,z)E_y(x,y,z)
\end{equation}
\begin{equation}
     \frac {\partial{ }} {\partial{x}} E_y(x,y,z)=-j\omega\mu_0\mu_r(x,y,z)H_z(x,y,z)
\end{equation}
\begin{equation}
     \frac {\partial{ }} {\partial{z}} E_y(x,y,z)=j\omega\mu_0\mu_r(x,y,z)H_x(x,y,z)
\end{equation}
Combining (4), (7) and (3), (6) together with (1), (2), results in following
\begin{equation}
\epsilon_r(x,-y,z)=\mu_r(x,y,z), \quad  \quad \quad \quad \quad\quad\quad\quad\quad \quad\quad\quad\quad\quad \mu_r(x,-y,z)=\epsilon_r(x,y,z)  
\end{equation}
Equation (9) which is followed from the 
dual of correspondence between the EM fields of the complementary pair describes the mirror reflection symmetries. By considering the above 
equations, the governing equation for the pseudospin up state can be written as
\begin{equation}
\begin{pmatrix}
0 & -\frac {\partial{ }} {\partial{z}} & 0 \\
-\frac {\partial{ }} {\partial{z}} & 0 & \frac {\partial{ }} {\partial{x}} \\
0 & \frac {\partial{ }} {\partial{x}} &0 

\end{pmatrix}
\psi^+(x,-y,z)=j\omega \sqrt{\epsilon_0\mu_0}\epsilon_r(x,-y,z)\psi^+(x,y,z)
\end{equation}

$
\psi^+(x,y,z)=(\sqrt{\epsilon_0}E_x(x,-y,z)-\sqrt{\mu_0}H_x(x,y,z) \quad \sqrt{\epsilon_0}E_y(x,-y,z)+\sqrt{\mu_0}H_y(x,y,z)\quad \sqrt{\epsilon_0}E_z(x,-y,z)-\sqrt{\mu_0}H_z(x,y,z))^T
$
\newline
\newline
where $\psi^+$
is referred to as the pseudospin up state. Note that, generally speaking, we have to 
consider all conditions. Hence, in a similar way, if the field components of the TE surface at
point $(x,y,z)$ are as $H_x, E_y$, and $H_z$ then the field components of the TM surface, i.e., $E_x, H_y,$ and 
$E_z$ at any arbitrary point $(x,-y,z)$ can be written as
\begin{equation}
H_y(x,-y,z)=-\sqrt{\frac {\epsilon_0} {\mu_0}} E_y(x,y,z)
\end{equation}
\begin{equation}
E_x,_z(x,-y,z)=\sqrt{\frac {\mu_0} {\epsilon_0}} \,H_x,_z(x,y,z)
\end{equation}
Equations (11), (12) represent the mutual relationship between EM field components of
pseudospin down state. According to the definition of the pseudospin down state, $H_y$
field 
component at any point $(x,-y,z)$ and 
$E_y$
field component at any point $(x,y,z)$ are out of phase, 
while the 
$H_x,_z$
field components and the 
$E_x,_z$
field components are in phase. Additionally, the 
linear combination of two surface modes with orthogonal polarizations propagating with the 
same momentum $\beta _{T_E}=\beta _{T_M}=\beta$ yields the pseudospin down state, whose governing equation can 
be written as
\begin{equation}
\begin{pmatrix}
0 & -\frac {\partial{ }} {\partial{z}} & 0 \\
-\frac {\partial{ }} {\partial{z}} & 0 & \frac {\partial{ }} {\partial{x}} \\
0 & \frac {\partial{ }} {\partial{x}} &0 
\end{pmatrix}
 \psi^-(x,-y,z)  =-j\omega \sqrt{\epsilon_0\mu_0}\epsilon_r(x,-y,z)\psi^-(x,y,z)
\end{equation}
$
\psi^-(x,y,z)=(\sqrt{\epsilon_0}E_x(x,-y,z)+\sqrt{\mu_0}H_x(x,y,z) \quad \sqrt{\epsilon_0}E_y(x,-y,z)-\sqrt{\mu_0}H_y(x,y,z)\quad \sqrt{\epsilon_0}E_z(x,-y,z)+\sqrt{\mu_0}H_z(x,y,z))^T
$
\newline
\newline
where $\psi^-$ is referred to as pseudospin down state. Thus the interface of two EM dual boundary 
conditions in the configuration of complementary metasurfaces, forms a pair of mirror images 
and supports symmetry-protected orthogonal states propagating in adverse directions and 
exhibit one-way propagation. These states are decoupled and they are associated with a time-reversal symmetry of the form,$\psi^+(x,y,z)=L_T\psi^-(x,y,z)$. Furthermore, these pseudospin 
states can be transformed to each other by EM duality of the form $\psi^+(x,y,z)=L_D\psi^-(x,-y,z)$, where $L_T$ and $L_D$ are the time-reversal and EM duality 
operators, respectively. According to equations (9), two complementary pairs satisfy the 
mirror reflection symmetries about the $x-z$ plane. This condition can be generalized to a 
universal state, thus it is possible to form pseudospin-polarized open and closed waveguides by 
establishing the inversion symmetries in a proper configuration.

\section*{Ultra Wide-Band One-dimensional Waveguide with Symmetric Line Mode}
As discussed earlier, there have been established conditions for the existence of LW and the yielded mode by characterizing the interfaced planes merely by one-layer complementary sheets. However, the LW suffers from asymmetric spatial field confinement in which higher field concentration is clearly seen on one side of complementary surfaces [4]. In this section, we aim to overcome this inherent limitation by using a conducting strip line of the width $ w< \lambda_0/15 $ and two pairs of complementary metasurfaces. Meanwhile, our complementary structures satisfy two main requirements of metasurfaces i.e. subwavelength thickness of array and subwavelength separation of the scatterer. For dispersive and isotropic complementary impedance surfaces characterizing the boundaries of our pseudospin polarized waveguides, the corresponding surface impedances for TM and TE
polarized waves can be written as
\begin{equation}
Z_s^{T^M}=j\eta_0/\zeta_{T_M}(\omega), \quad  \quad \quad \quad \quad\quad\quad\quad\quad \quad\quad\quad\quad\quad Z_s^{T^E}=-j\eta_0\times \zeta_{T_E}(\omega)
\end{equation}
where $\eta_0$ denotes the intrinsic impedance of free space and $\zeta$ is a dimensionless real parameter
that is both frequency and spatially dispersive. In general, both impedances are complex valued, 
but for non-resonant metasurfaces in the microwave frequencies due to the finite conductivity of 
metal, dissipation losses are negligible [26]. An outcome of Babienet's principle affirms that
\begin{equation}
    Z_s^{T^E}\times Z_s^{T^M}=\frac {\eta_0^2} {4}
\end{equation}
Substitution of (14) in (15) yields
\begin{equation}
 \zeta_{T_M}(\omega)=4\times  \zeta_{T_E}(\omega) 
\end{equation}
Two impedance surfaces with opposite/complementary electromagnetic responses whose $\zeta$ satisfy (16) are dual EM metasurfaces. This condition corresponds to complementary
impedance boundaries supporting TM and TE surface waves with identical wavenumbers in a 
manner that they show equal response to EM fields and preserve the mirror reflection 
symmetries.\\
\begin{figure}[ht]
\centering
\includegraphics[width=\linewidth]{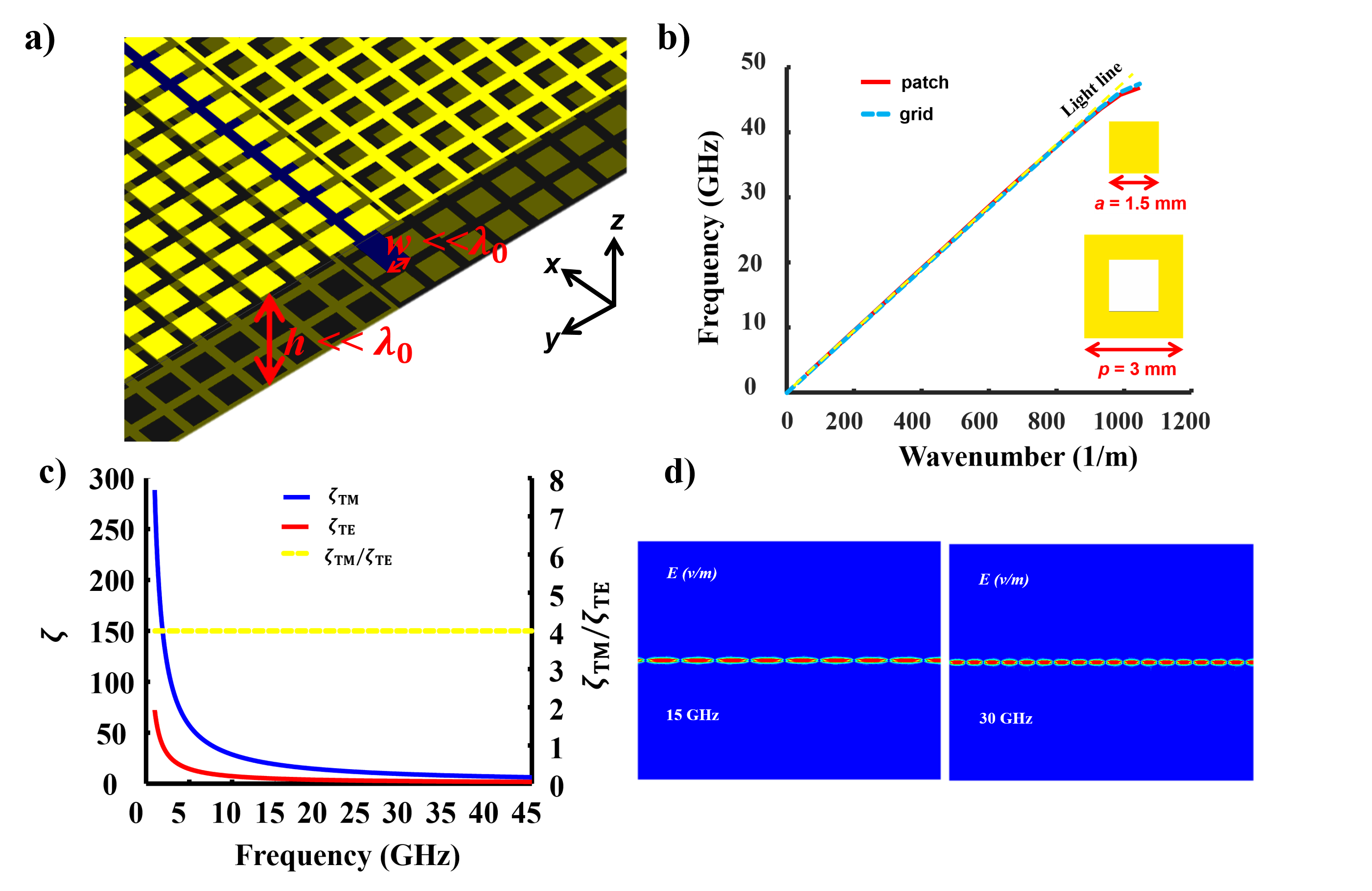}
\caption{  (a) Schematic of a pseudospin-polarized one-dimensional waveguide, (b) dispersion characteristics of the TE and TM surfaces with geometrical parameters, (c) $\zeta$ values of the corresponding complementary metasurfaces, and (d) electric field distribution of LW at different frequencies.}
\label{fig:Figure2}
\end{figure}
The theorem of pseudospin states can be applied to establish new waveguiding 
phenomena based on the spin-filtered characteristic that, in turn, permits manipulating EM 
energy in a scattering-free way. Here, a new waveguide presented that supports a symmetric LW. The proposed pseudospin-suppressed waveguide provides extremely wide working bandwidth and better performance along bend paths. The waveguide consisting of a strip line sandwiched between four impedance surfaces with complementary EM responses, as shown in Figure 2(a). The constituent impedance boundaries satisfying $\epsilon_r(x,y,z)=\mu_r(x,y,-z)$ and $\mu_r(x,y,z)=\epsilon_r(x,y,-z)$ 
symmetries. Hence, a system of robust spin states is formed whose pseudospin polarizations are exclusively defined by the direction of wavevector. Note that, in the proposed waveguide the strip is on the 
x-y plane. Hence, such a circumstance does not violate the condition of preserving the mirror 
reflection symmetries. It should be noted that, in general, the strip can be replaced by coplanar 
strips, spoof SPP structures, two interfaced complementary metasurfaces, and so on. The geometry of the proposed structure was characterized by the
width of the strip (\emph{w}) and separation distance between two complementary metasurfaces (\emph{h}). Both of the parameters have subwavelength values; consequently, the waveguide supports the tightly confined line mode, which is bounded to the strip line. The space inside the waveguide filled with \emph{ Teflon} with $\epsilon_r=2.1, \delta_t=0.001$. Note that, as long as the value of \emph{h} remains in the subwavelength limit, i.e., $\emph{h}<<\lambda$, the EM duality
requirement and, hence the mirror reflection symmetries are satisfied adequately well [24]. Figure 
2(b) depicts the unit cell design of one-layer complementary metasurfaces and corresponding dispersion relation for the TM and TE surface waves. The dispersion curves of the freestanding 
complementary structures are nearly the same and they overlap over a wide frequency range. In 
addition, the corresponding plot of $\zeta$ versus frequency is presented in Figure 2(c). The
existence of inequality in $\zeta$ for fully complementary surfaces forces the 
higher concentration of EM fields toward the surface with higher $\zeta$. This asymmetric behavior especially in case of the one-layer waveguide leads to increasing scattered-bend 
losses. Conversely, since the supporting complementary surfaces do not have a bandgap, hence EM energy can be scattered into the surrounding area. However, exploiting a bandgap structure yields a reduction in operating bandwidth [5]. This can be alleviated by depositing a second layer, which is complementary to the first sheet. This condition results in the equality of $\zeta$ about the x-z plane; hence, EM energy confines symmetrically to the interface line, as shown in Figure 2(d). In addition, the presence of the strip line leads to improving the performance of the waveguide along bent paths.\\
\begin{figure}[ht]
\centering
\includegraphics[width=\linewidth]{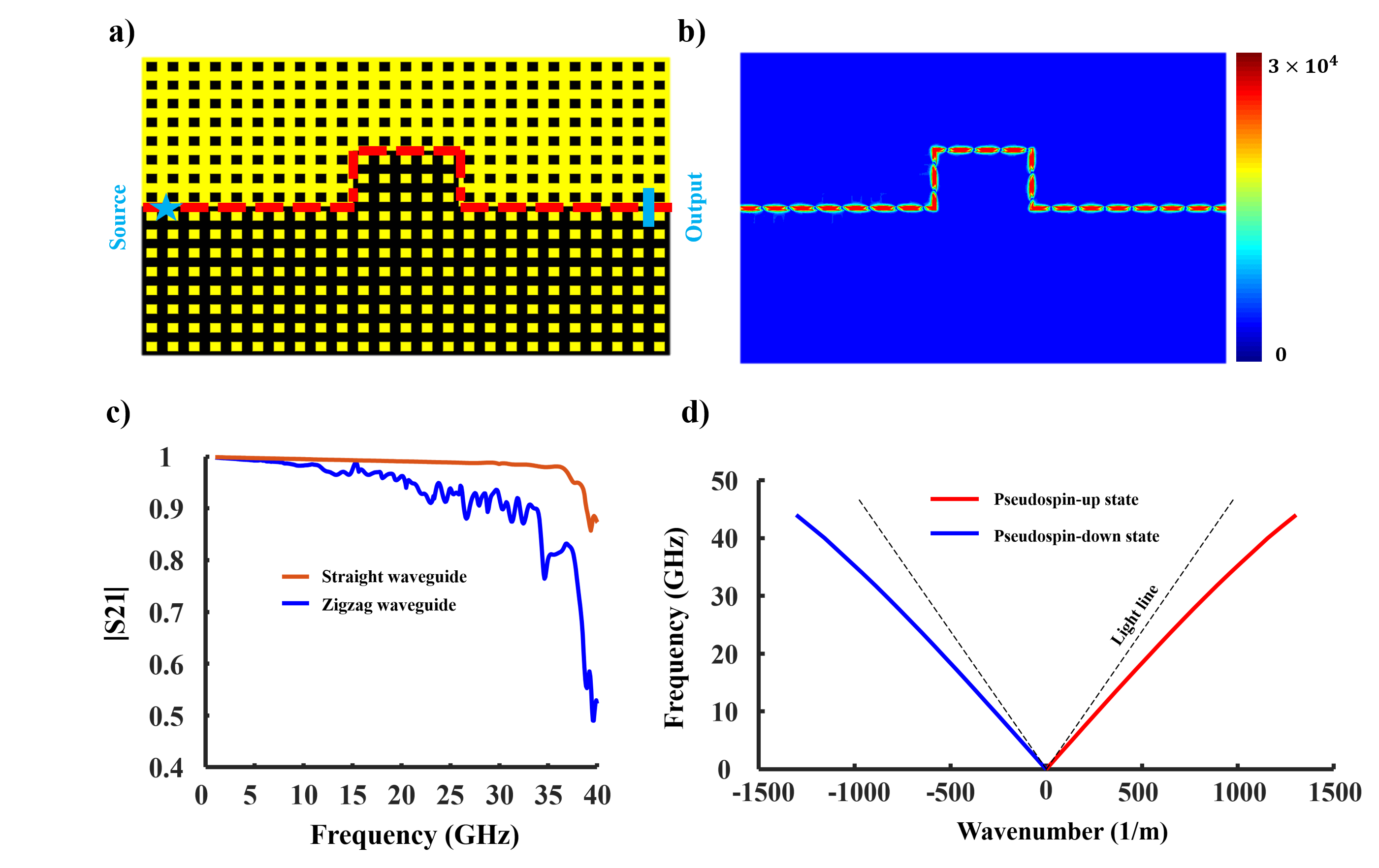}
\caption{ (a) The schematic of a sharp bend waveguide with multiple bends, (b ) propagation of the hybrid spin state along sharp bends, (c) the simulated transmittance, indicating low backscattering losses and extreme bandwidth, and (d) dispersion characteristic of LW.}
\label{fig:Figure3}
\end{figure}
Our pseudospin-polarized systems provide wideband operating bandwidth over conventional 
waveguides, as they are safe from the bandwidth restriction connected by band gaps in photonic 
crystals. Indeed, as long as the supporting metasurfaces share the same dispersion over a wide 
frequency range the constructive interaction between the reciprocal counterparts of EM fields, 
yields a pair of hybrid spin states exhibiting ultra-wide EM response, with robustness to time-reversal symmetry conserved disorders [3,27]. To analyze the ultra-wideband behavior of LW, we introduce a zigzag bend operating as a waveguide with multiple bends of the angles 90-degree. The schematic of the zigzag pathway is shown in Figure 3(a). Note that, the robustness of pseudospin states against sharp bends preserved by the boundary-mirror symmetries as long as the scatterer do not result in spin-flip transition. Figure 3(b) shows the propagation of the pseudospin-polarized LW along the bent path proving the backscattering immune spin-dependent transmission. On the other hand, as shown in Figure 3(c), a transmittance ($\lvert S21 \rvert$) of about 99\% is obtained in the frequency band of 1-38 GHz for the straight waveguide. Meanwhile, the bandwidth of the waveguide is not completely immune to perturbations so that the transmittance of about 95\% is achieved in the frequency range of 1-34 GHz. It should be noted that there is an increase of losses in proportion to increasing frequency; however, such perturbations cannot scatter the pseudospin in reverse direction or break spin degeneracy. Figure 3(d) plots the dispersion diagram of LW. The line mode forms the pseudospin-up state with wavenumber +$k$ and its time-reversal pair, the pseudospin-down state with wavenumber -$k$. The dispersion curve has been calculated for a suspended structure using the eigenmode solver of COMSOL Multiphysic.  
\section*{Ultra Wide-Band Pseudospin-polarized Slotline Waveguide}
Next, we investigate another possible realization of an open boundary pseudospin-polarized 
system which can be implemented by establishing EM duality conditions. Here, a new slotline
waveguide with multi-octave bandwidths has been demonstrated which provides the 
possibility to suppress backward losses. Consider a classic slotline waveguide bounded by two 
parallel complementary boundaries, preserving the spatial inversion symmetries of the form $\epsilon_r(x,y,z)=\mu_r(x,y,-z)$ and $\mu_r(x,y,z)=\epsilon_r(x,y,-z)$, as shown in Figure 4(a). Assuming that 
the conducting ground planes surrounded by air. Alternatively, they can be centered in a dielectric material with $\epsilon(r)\approx\mu(r)$ characteristic, which satisfies the condition of preserving the mirror
symmetries at any given point. Additionally, such a medium can be used as a substrate or a superstrate in pseudospin polarized waveguides with arbitrary configurations. Figure 4(b) presents the dispersion characteristics and the 
geometric illustration of the complementary metasurfaces. The unit cell period \emph{p} and the 
geometric parameter of the unit cell \emph{a} are set as \emph{p}=1.5mm and \emph{a}= 0.6,1, 1.3 mm. The coincidence 
of the dispersion curves over a wide frequency range warranted by EM duality. 
Therefore, the mutual interaction between two surface modes of the same momentum gives rise 
to wideband pseudospin states.
\begin{figure}[ht]
\centering
\includegraphics[width=\linewidth]{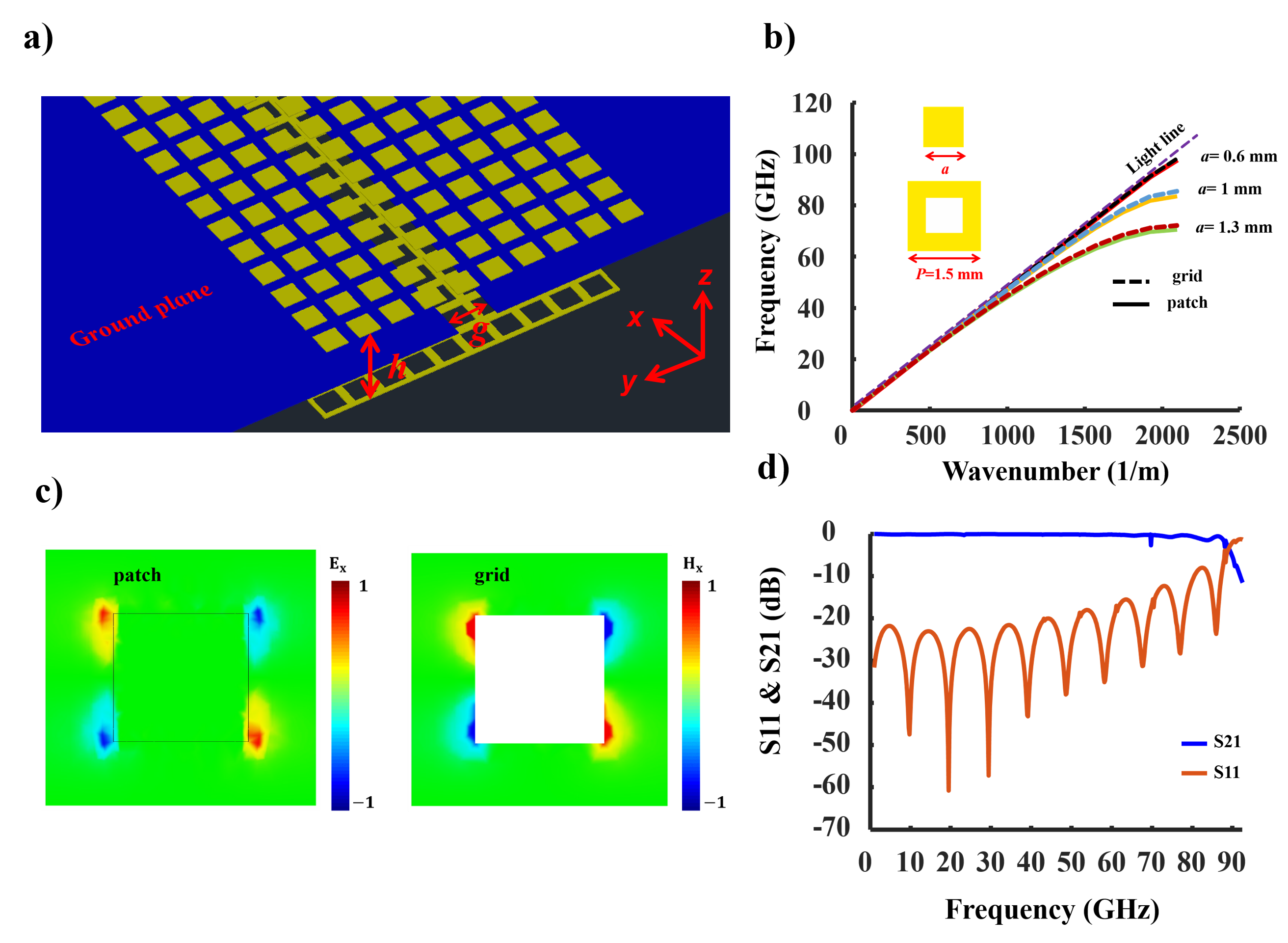}
\caption{  (a) A schematic representation of the proposed slotline waveguide, where the structural parameters are g=0.5mm, and d=0.4mm, (b) dispersion curves of two subwavelength complementary sheets (c) the normalized near-field distribution in complementary-unit-cells, and (d) transmission and reflection coefficients of the proposed waveguide obtained for \emph{a}=0.6 . }
\label{fig:Figure4}
\end{figure}

 Figure 4(c) depicts the reciprocal relationship between the x 
components of the E and H fields of the supporting complementary metasurfaces. This 
relationship is a direct consequence of EM duality that, in turn, results in the formation 
of the pseudospin states in a system subject to EM duality characteristics. Figure 4(d) shows the 
simulated scattering parameters reflection coefficient ($S11$) and, isolation ($S21$), of the 
proposed configuration. The slot line waveguide has a mean value of isolation of - 0.13 dB over a broad frequency range of 1-68 GHz. The resulting parameters, obtained bya full wave simulation using ANSYS HFSS, which is a FEM-based commercial software.\\
\begin{figure}[ht]
\centering
\includegraphics[width=\linewidth]{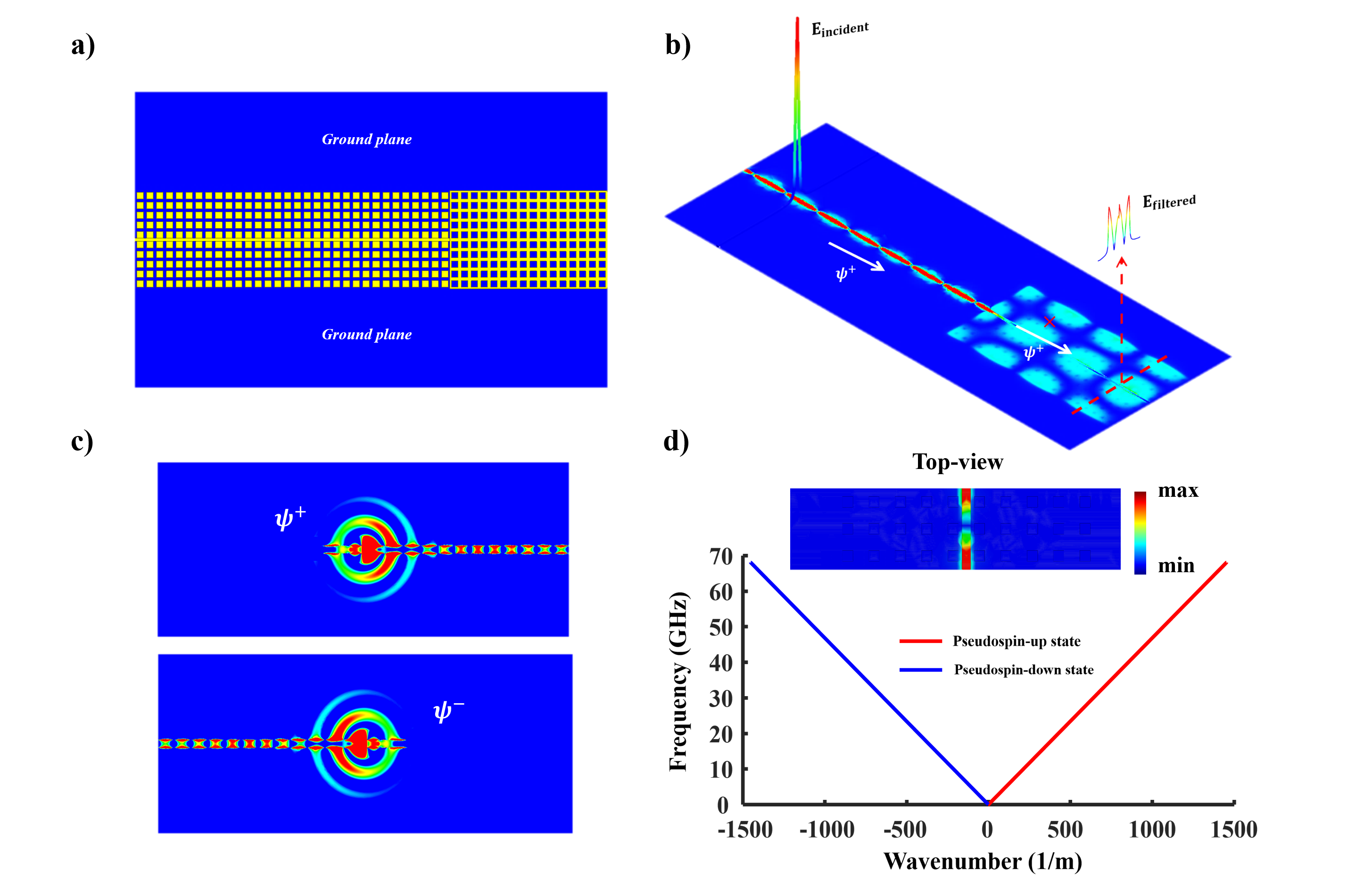}
\caption{(a) A schematic of deformed waveguide, where the boundary conditions are reversed in the structure, (b) spin-locking of a TEM mode at a pseudospin polarized open waveguide with complementary impedance boundaries. The obtained incident and filtered field profiles manifest that the structure is a spin-filtered waveguide, (c) unidirectional excitation of the pseudospin states proving polarization-momentum locking feature of the states, and (d) the dispersion diagram and the mode distribution of the slotline waveguide.}
\label{fig:Figure5}
\end{figure}
One crucial feature of the proposed slot line waveguide is that it supports momentum-spin locked states. This feature is also inherent to every fast decaying wave with evanescent chracteristic, in the transverse plane to the wave momentum [28]. In addition, this universal property, which obliges the handedness of waves to be defined uniquely by the transport direction, has been observed at surface plasmon polaritons [29]. To demonstrate the directionality of the pseudospin states, as shown in Figure 5(a), we change the boundary conditions in the proposed structure. Obviously, the time-reversal invariant waveguide prevents the propagation of $\psi^+$ along the defect region, as shown in Figure 5(b). Hence, the system is a spin-suppressed route so that the definition of the pseudospin states tied to the direction of the momentum. Alternatively, the intensity profiles for the incident and filtered electric fields confirm that robust spin states are locked to the wavevector. This result agrees well with our predicted theory that enables forming pseudospin-filtered systems with unique characteristics. Figure 5(c) shows the unidirectional excitation of the pseudospin states using a chiral point source carrying the same spin angular momentum as that of the slot line waveguide mode [30-32]. The chiral source placed at the center of the waveguide, consisting of two out of phase electric point sources with orthogonal configuration. The optional excitation of the spin states demonstrates that the one-way pseudospin is strictly characterized by the direction of propagation. In the same manner, the pseudospin states can be excited via a pair of dual point sources (i.e., electric and magnetic Hertzian dipoles) with specific phase relations [3]. The dispersion relation of the proposed slotline waveguide and mode distribution of a supercell is presented in Figure 5(d). The boundaries of the supercell along the propagation direction are applied with a periodic boundary while the other borders are chosen as radiation boundary conditions. The propagation constant of the fundamental mode is the same as the TEM mode, which has a linear dispersion relation. 
 \section*{ Psedospin-Polarized Closed Waveguide }
Open waveguides are characterized by imperfectly reflecting or open boundaries, while closed 
waveguides are described by being entirely enclosed with highly reflecting walls. In ordinary 
closed waveguides, EM energy is confined inside the boundaries considered as either 
perfectly electric walls with zero impedances or perfectly magnetic walls with zero admittances [34]. On the other hand, a pseudospin-polarized closed waveguide has been formed by properly 
preserving the mirror reflection symmetries using two pairs of perfectly reflecting electric and 
magnetic boundaries, in which case EM duality is clearly established between the walls. 
Nonetheless, PMCs are generally realized with Sievenpiper high-impedance surfaces [35], which have a limited bandwidth. On the other hand, \emph{Chen et al} propose the implementation of an effective PMC 
boundary using periodic PEC structures, which have a non-resonant nature [13]. This method introduces additional mirror symmetries. However, it is unfeasible to implement a closed
pseudospin-polarized waveguide using periodic PEC boundary conditions. Alternatively, we propose the use of non-resonant 
complementary impedance boundaries which satisfy (16). 
\begin{figure}[ht]
\centering
\includegraphics[width=\linewidth]{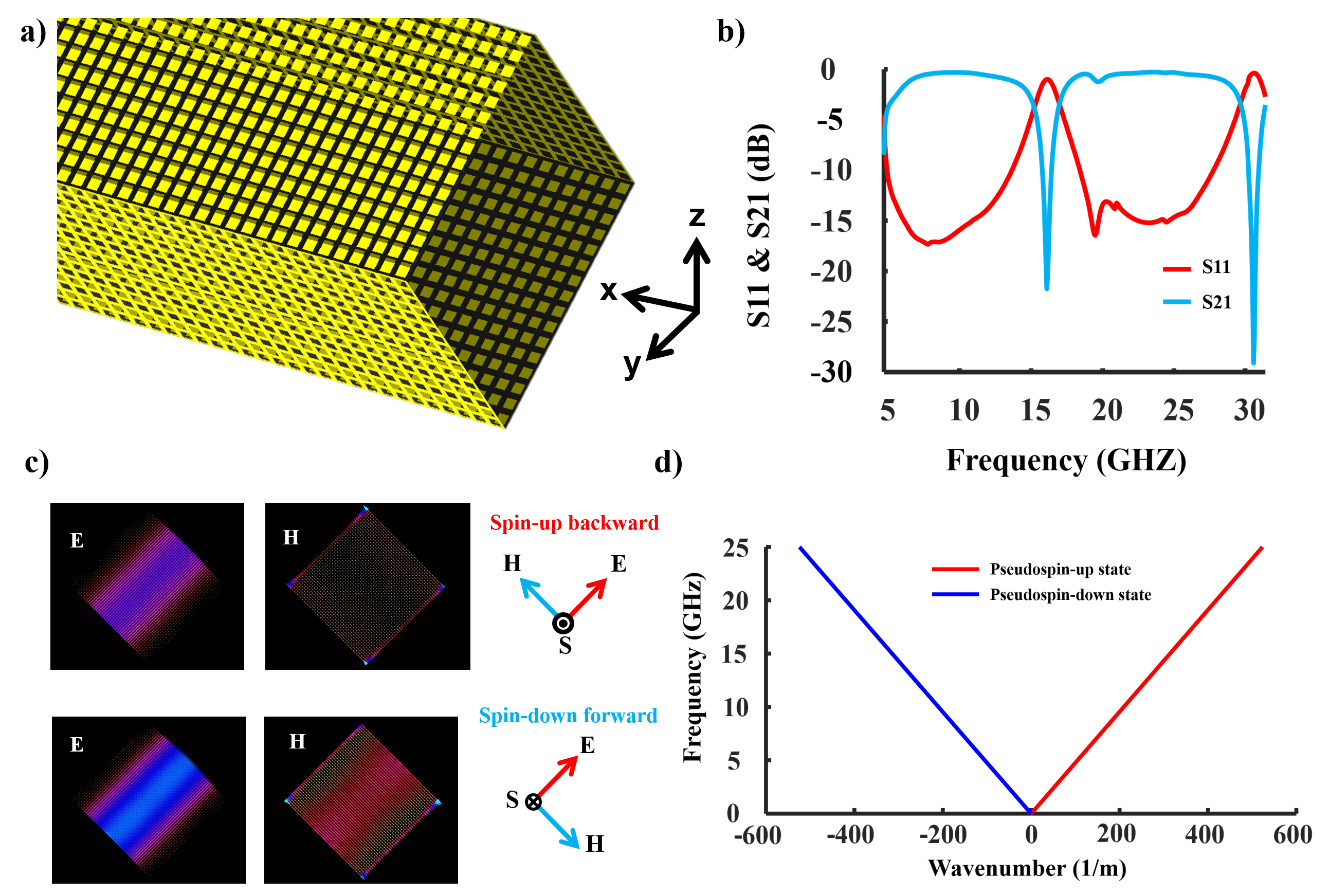}
\caption{(a) A schematic of the proposed closed waveguide. The side length of the waveguide consisting of 16 unit cells of the complementary metasurfaces, with the unit cell period of 2mm, and the gap spacing between adjoining patches of 0.9mm, (b) scattering parameters of the corresponding waveguide, (c) eigenvector fields distribution of two decoupled spin states over a cross section of the closed waveguide, and (d) dispersion diagram for the closed waveguide with complementary boundaries.}
\label{fig:Figure6}
\end{figure}

\begin{figure}[ht]
\centering
\includegraphics[width=\linewidth]{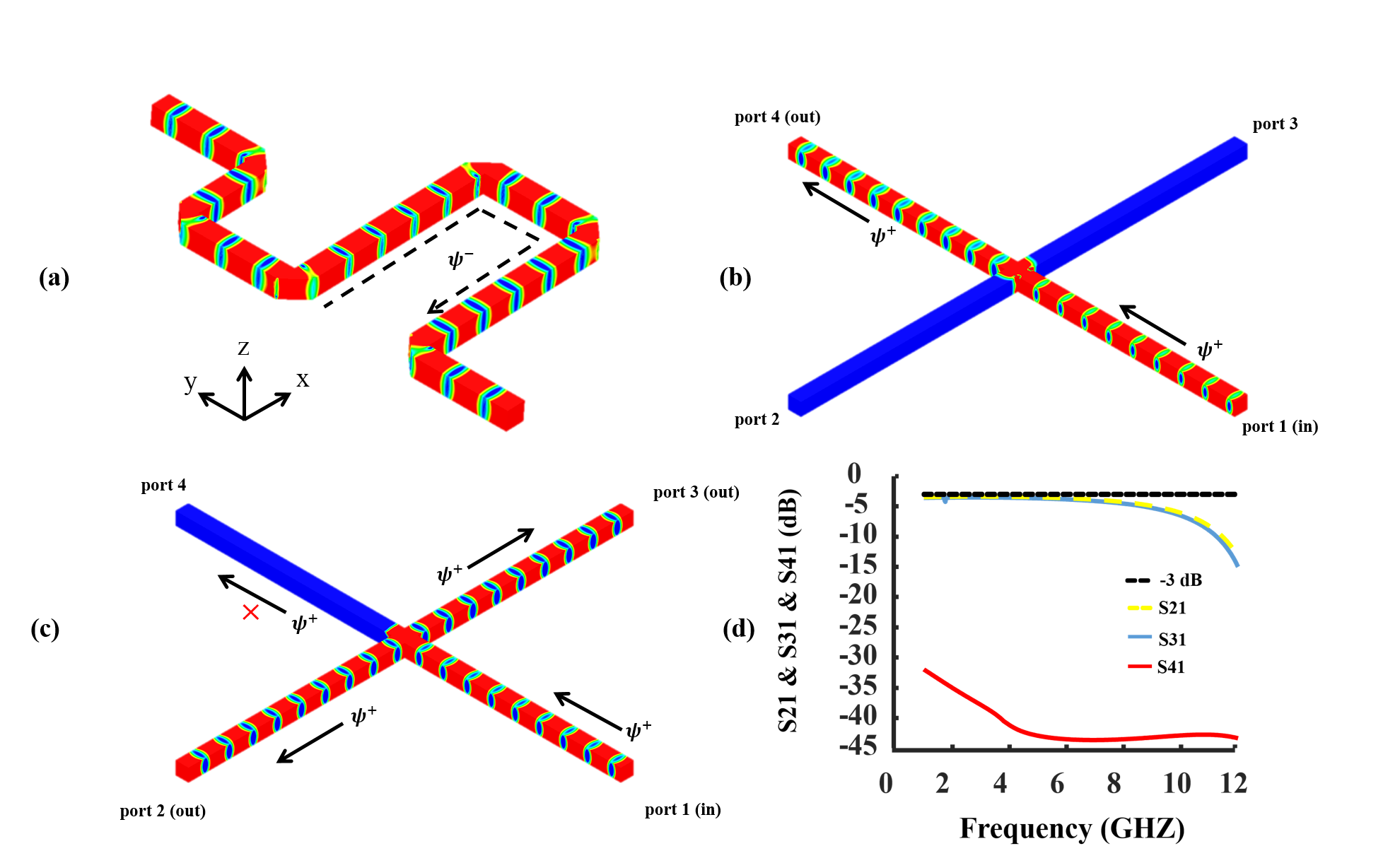}
\caption{(a) Guiding spin down state along multiple bends showing back scattering immune transportation, (b) energy transmission in a T-junction network, (c) the T-junction as a splitter based on spin-filtering feature, and (d) scattering parameters of the corresponding splitter structure.}
\label{fig:Figure7}
\end{figure}
\begin{figure}[ht]
\centering
\includegraphics[width=\linewidth]{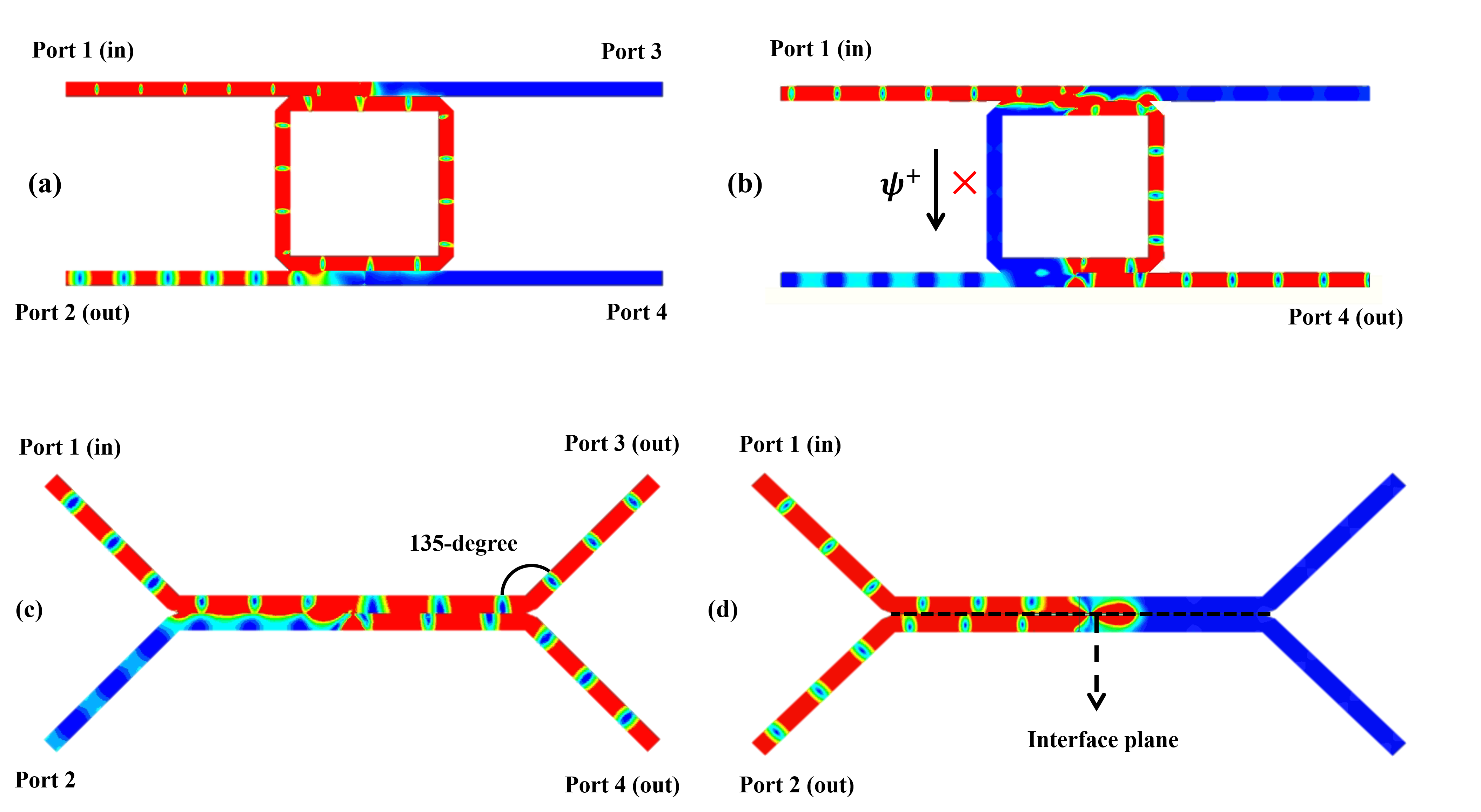}
\caption{ An illustration of potential applications of the spin-polarized closed waveguide with dynamic impedance boundary conditions: (a), and (b) guiding EM energy in an add-drop filter through dynamic paths, (c) a hybrid directional coupler, and (d) an unconventional guiding of EM energy in a directional coupler with complementry impedance walls.}
\label{fig:Figure8}
\end{figure}

Figure 6(a) shows a schematic of the proposed spin-polarized closed waveguide where the surface impedances or surface admittances 
of the walls relating the electric (E) and magnetic (H) field components tangential to the wall,
are both frequency and spatially dispersive. Note that the impedance of walls 
can be anisotropic, which is described by a tensor, rather than scalar surface impedance. We assume that the waveguide filled 
homogeneously with isotropic $Foam$ with $\epsilon=1.4, \mu=1$. Therefore, the nearly matched $\epsilon-\mu$ condition provides a very good approximation of a time-reversal invariant system. However, this 
condition only applies to the pseudospin waveguides with dual boundary conditions. Hence,
regarding the coordinate system in Figure 6(a), the proposed closed waveguide is mirror-symmetric about the x-y plane. Figure 6(b), shows a simulation plot of the scattering parameters 
$S11$ and $S21$. The fundamental mode is wideband in the 
frequency range 7.5-11.8 GHz with a $S21$ value of about -0.5dB. The vector eigenfields distributions of the pseudospin states over 
the cross-section of the waveguide are presented in Figure 6(c). There are two plane-polarized waves (i.e., 
spin-up backward and spin-down forward states) that essentially determined by the direction of 
propagation. Hence, the system forms a decoupled time-reversal pair of the pseudospin states, 
which are associated with EM duality. Figure 6(d) plots the dispersion relation of the
fundamental pseudospin states. As expected, the pseudospin TEM mode is the mode of 
propagation as the E and H fields are perpendicular to each other and to the direction of 
propagation.\\
The EM dual boundary conditions about the mirror plane, provide our closed 
waveguide with a backscattering-immune transportation feature. Here, various types of
perturbations are considered to show that the pseudospin configurations directly characterize the 
direction of momentum. For simplicity, we consider non-dispersive surface impedance walls, 
which means that the value of impedances is independent of the tangential wavenumber of the 
incident wave and frequency. Consider the case where the hybrid polarization $\psi^-$ propagates 
along consecutive bends, as shown in Figure 7(a). To satisfy the boundary continuity, there is no 
way to excite $\psi^+$ polarization. Since, the scatterer do not flip the pseudospin polarization, protection in the system takes effect and the mirror reflection symmetries preserve the 
propagating without remarkable reflections. Moreover, to show the spin-filtered 
characteristic, a T-junction structure is designed, as shown in figure 7(b). The excited pseudospin up state at port 1 couples into port 4, while the inverted boundary 
conditions along ports 2 and 3 prevent the coupling of $\psi^+$ into $\psi^-$. On the other hand, it is possible to form a -3dB coupler by changing the boundary conditions. As shown in Figure 7(c), EM power splits into ports 2 and 3 while scattering into port 4 is impossible. Consequently, the pseudospin states are merely defined according to the direction of propagation and the closed waveguide with complementary impedance walls demonstrates a spin-filtered route. Alternatively, as shown in Figure 7(d), the scattering parameters S21 and S31 of about -3dB and S41 below -30dB are obtained over a wide frequency range. Hence, the propagation is completely filtered along port 4. These results have been obtained by a driven mode analysis of HFSS for $Z_{T_E}=-j2\eta_0 , Z_{T_M}=j\eta_0/8$ and with a side length of the waveguide of 6mm.\\
The associated momentum-spin locking property can be exploited to form versatile network devices. Furthermore, it is possible to incorporate impedance surfaces with dual capacitive-inductive nature thus allowing for forming unconventional passive circuits with dynamic configurations [30,36]. Especially, this includes graphene-based metasurfaces showing great potential applications in the terahertz and optical regimes [37-39]. As an illustration, Figure 8(a), and 8(b) shows an add-drop filter design that enables transforming EM energy along arbitrary paths simply by switching a capacitive/inductive impedance to inductive/capacitive impedance. On the other hand, Figure 8(c) shows a spin-filtered-based directional coupler that can couple EM power into any desired branch. The structure is consisting of two closed waveguides with complementary impedance walls, which are stuck together while a rectangular hole at the interface plane is created. The coupler works based on the spin-filtering feature, which allows for manipulating EM waves inside the closed system. This coupler can split power equally between two output ports like a hybrid coupler. However, unlike traditional directional couplers which couple EM energy in one direction, our structure can couple energy in any desired direction [39]. For instance, reversing boundaries along ports 3 and 4 prevents the propagation through these paths. Hence, EM energy rotates 180-degree and couples to port 2, as shown in Figure 8(d).
\newline
\newline
\newline
\newline
\newline
\newline
\newline
\newline
\newline
\newline
\newline
\section{conclusion}
In summary, we have shown how ultra-wideband pseudospin-dependent transport can be achieved simply using 2D patterned and complementary sheets. Furthermore, we implemented the theorem of pseudospin states on several open and closed waveguides. The findings change our point of view of the classic waveguides such that the concept of all waveguides can be generalized to wideband spin-polarized waveguides as well as our adopted theory enables designing new one-way ultra wideband systems with robustness to perturbations.

\section*{Refrences}

1.	Olyslager F. Electromagnetic waveguides and transmission lines. OUP Oxford; 1999 May 27.\\
2.	Rozzi T, Mongiardo M. Open electromagnetic waveguides. IEE Electromagnetic waves series. 1997;43.\\
3.	Dia’aaldin JB, Sievenpiper DF. Guiding waves along an infinitesimal line between impedance surfaces. Physical review letters. 2017 Sep 8;119(10):106802.\\
4.	Zafari K, Oraizi H. Surface Waveguide and Y Splitter Enabled by Complementary Impedance Surfaces. Physical Review Applied. 2020 Jun 11;13(6):064025.\\
5.	Bisharat DA, Sievenpiper DF. Electromagnetic‐Dual Metasurfaces for Topological States along a 1D Interface. Laser \& Photonics Reviews. 2019 Oct;13(10):1900126.\\
6.	Li A, Singh S, Sievenpiper D. Metasurfaces and their applications. Nanophotonics. 2018 Jun 1;7(6):989-1011.\\
7.	Liu Y, Ouyang C, Zheng P, Ma J, Xu Q, Su X, Li Y, Tian Z, Gu J, Liu L, Han J. Simultaneous Manipulation of Electric and Magnetic Surface Waves by Topological Hyperbolic Metasurfaces. ACS Applied Electronic Materials. 2021 Sep 15;3(9):4203\\
8.	Glybovski SB, Tretyakov SA, Belov PA, Kivshar YS, Simovski CR. Metasurfaces: From microwaves to visible. Physics reports. 2016 May 24;634:1-72.\\
9.	Hsiao HH, Chu CH, Tsai DP. Fundamentals and applications of metasurfaces. Small Methods. 2017 Apr;1(4):1600064.\\
10.	Kildishev AV, Boltasseva A, Shalaev VM. Planar photonics with metasurfaces. Science. 2013 Mar 15;339(6125).\\
11.	Yu N, Capasso F. Flat optics with designer metasurfaces. Nature materials. 2014 Feb;13(2):139-50.\\
12.	Meinzer N, Barnes WL, Hooper IR. Plasmonic meta-atoms and metasurfaces. Nature Photonics. 2014 Dec;8(12):889-98.\\
13.	Chen WJ, Zhang ZQ, Dong JW, Chan CT. Symmetry-protected transport in a pseudospin-polarized waveguide. Nature communications. 2015 Sep 23;6(1):1-8.\\
14.	Khanikaev AB, Mousavi SH, Tse WK, Kargarian M, MacDonald AH, Shvets G. Photonic topological insulators. Nature materials. 2013 Mar;12(3):233-9.\\
15.	Chen WJ, Jiang SJ, Chen XD, Zhu B, Zhou L, Dong JW, Chan CT. Experimental realization of photonic topological insulator in a uniaxial metacrystal waveguide. Nature communications. 2014 Dec 17;5(1):1-7.\\
16.	Lu L, Joannopoulos JD, Soljačić M. Topological photonics. Nature photonics. 2014 Nov;8(11):821-9.\\
17.	Slobozhanyuk A, Mousavi SH, Ni X, Smirnova D, Kivshar YS, Khanikaev AB. Three-dimensional all-dielectric photonic topological insulator. Nature Photonics. 2017 Feb;11(2):130-6.\\
18.	Kang Y, Ni X, Cheng X, Khanikaev AB, Genack AZ. Pseudo-spin–valley coupled edge states in a photonic topological insulator. Nature communications. 2018 Aug 2;9(1):1-7.\\
19.	Zhou P, Liu GG, Ren X, Yang Y, Xue H, Bi L, Deng L, Chong Y, Zhang B. Photonic amorphous topological insulator. Light: Science \& Applications. 2020 Jul 24;9(1):1-8.\\
20.	Wang Z, Chong Y, Joannopoulos JD, Soljačić M. Observation of unidirectional backscattering-immune topological electromagnetic states. Nature. 2009 Oct;461(7265):772-5.\\
21.	Bliokh KY, Smirnova D, Nori F. Quantum spin Hall effect of light. Science. 2015 Jun 26;348(6242):1448-51.\\
22.	Ao X, Lin Z, Chan CT. One-way edge mode in a magneto-optical honeycomb photonic crystal. Physical Review B. 2009 Jul 15;80(3):033105.\\
23.	Dong JW, Chen XD, Zhu H, Wang Y, Zhang X. Valley photonic crystals for control of spin and topology. Nature materials. 2017 Mar;16(3):298-302.\\
24.	Liang C, Liu L, Chai J, Xiang H, Han D. Duality of spoof surface plasmon polaritons on the complementary structures of ultrathin metal films. Annalen der Physik. 2019 Nov;531(11):1900138.\\
25.	Zentgraf T, Meyrath TP, Seidel A, Kaiser S, Giessen H, Rockstuhl C, Lederer F. Babinet’s principle for optical frequency metamaterials and nanoantennas. Physical Review B. 2007 Jul 10;76(3):033407.\\
26.	Ma X, Mirmoosa MS, Tretyakov SA. Parallel-plate waveguides formed by penetrable metasurfaces. IEEE Transactions on Antennas and Propagation. 2019 Aug 16;68(3):1773-85.\\
27.	Jung J, Park H, Park J, Chang T, Shin J. Broadband metamaterials and metasurfaces: a review from the perspectives of materials and devices. Nanophotonics. 2020 Sep 2;9(10):3165-96.\\
28.	Van Mechelen T, Jacob Z. Universal spin-momentum locking of evanescent waves. Optica. 2016 Feb 20;3(2):118-26.\\
29.	Lin J, Mueller JB, Wang Q, Yuan G, Antoniou N, Yuan XC, Capasso F. Polarization-controlled tunable directional coupling of surface plasmon polaritons. Science. 2013 Apr 19;340(6130):331-4.\\
30.	Dia’aaldin JB, Sievenpiper DF. Manipulating line waves in flat graphene for agile terahertz applications. Nanophotonics. 2018 May 1;7(5):893-903.\\
31.	Bliokh KY, Rodríguez-Fortuño FJ, Nori F, Zayats AV. Spin–orbit interactions of light. Nature Photonics. 2015 Dec;9(12):796-808.\\
32.	Rodríguez-Fortuño FJ, Marino G, Ginzburg P, O’Connor D, Martínez A, Wurtz GA, Zayats AV. Near-field interference for the unidirectional excitation of electromagnetic guided modes. Science. 2013 Apr 19;340(6130):328-30.\\
33.	Mahmoud SF. Electromagnetic waveguides: theory and applications. IET; 1991.\\
34.	Sievenpiper D, Zhang L, Broas RF, Alexopolous NG, Yablonovitch E. High-impedance electromagnetic surfaces with a forbidden frequency band. IEEE Transactions on Microwave Theory and techniques. 1999 Nov;47(11):2059-74.\\
35.	Padooru YR, Yakovlev AB, Kaipa CS, Hanson GW, Medina F, Mesa F. Dual capacitive-inductive nature of periodic graphene patches: Transmission characteristics at low-terahertz frequencies. Physical Review B. 2013 Mar 4;87(11):115401.\\
36.	Fei Z, Rodin AS, Andreev GO, Bao W, McLeod AS, Wagner M, Zhang LM, Zhao Z, Thiemens M, Dominguez G, Fogler MM. Gate-tuning of graphene plasmons revealed by infrared nano-imaging. Nature. 2012 Jul;487(7405):82-5.\\
37.	Wang M, Zeng Q, Deng L, Feng B, Xu P. Multifunctional graphene metasurface to generate and steer vortex waves. Nanoscale research letters. 2019 Dec;14(1):1-7.\\
38.	Zhang Y, Li T, Chen Q, Zhang H, O’Hara JF, Abele E, Taylor AJ, Chen HT, Azad AK. Independently tunable dual-band perfect absorber based on graphene at mid-infrared frequencies. Scientific reports. 2015 Dec 22;5(1):1-8.\\
39.	Wang Q, Pirro P, Verba R, Slavin A, Hillebrands B, Chumak AV. Reconfigurable nanoscale spin-wave directional coupler. Science advances. 2018 Jan 1;4(1):e1701517.
\end{document}